\documentclass[referee]{raa}           
\usepackage{graphicx,times}
\usepackage{amssymb,amsmath}
\usepackage[colorlinks]{hyperref}
\hypersetup{colorlinks,citecolor=blue,linkcolor=blue,urlcolor=blue}  
\begin{document}
\title{The non-Gaussian distribution of galaxies gravitational fields}
 \volnopage{Vol.0 (200x) No.0, 000--000}      
 \setcounter{page}{1}   
 \author{Vladimir Stephanovich\inst{}, W{\l}odzimierz God{\l}owski\inst{}}
 \institute{ Uniwersytet Opolski, Institute of Physics, ul.  Oleska  48,
45-052 Opole, Poland; {\it stef@uni.opole.pl}\\ 
\vs \no
   {\small Received; accepted}
}
\abstract{We perform a theoretical analysis of the observational dependence between angular momentum of the 
galaxy clusters and their mass (richness), based on the method introduced in our previous paper. 
For that we obtain the distribution function of astronomical 
objects (like galaxies and/or smooth halos of different kinds) gravitational fields due to their 
tidal interaction. Within the statistical method of Chandrasekhar we are able to show that the 
distribution function is determined by the form of interaction between objects and for multipole 
(tidal) interaction it is never Gaussian. Our calculation permits to demonstrate how the alignment 
of galaxies angular momenta depend on the cluster richness. The specific form of the 
corresponding dependence is due to assumptions made about cluster morphology. Our approach 
also predicts the time evolution of stellar objects angular momenta within CDM and $\Lambda$CDM models.
Namely, we have shown that angular momentum of galaxies increases with time.}
\keywords{Galaxy:General --- Galaxy:Formation
\authorrunning{Stephanovich, God{\l}owski}            
\titlerunning{Non-Gaussian distribution}  
\maketitle

\section{Introduction}

In the models of galaxies and their structures formation the distribution of gravitational fields 
of their constituents play the decisive role. Many scenarios of such formation have been around 
for some time (\cite{peb69,Sunyaew72,Zeldovich70,Doroshkevich73,Shandarin74,Dekel85, Silk83}). 
Under the influence of new observational data, these scenarios are constantly being revised and 
improved, see (\cite{Shandarin12,Giah14}) and references therein for latest discussion. The  
main controversy here is how galaxies acquire their angular moments, which render subsequently 
to those of galaxy clusters and larger structures. On the other hand, this angular moment 
acquisition is intimately related to the above gravitational fields distribution.

The commonly accepted model of the Universe is spatially flat homogeneous and isotropic 
$\Lambda$CDM model. The galaxy clusters in this model are formed as a result of adiabatic and 
almost scale invariant Gaussian fluctuations (\cite{Silk68,Peebles70,Sunyaew70}). This assumption 
is the base of the so-called hierarchical clustering model (\cite{dor70,Dekel85,peb69}), the most 
popular scenario of galaxies formation. Note, however, the presence of the models with 
non-Gaussian initial fluctuations, see (\cite{bart04}) and references therein. This non-Gaussianity, 
however, has been postulated in certain form rather then calculated. 
At the same time, the non-Gaussian distributions can be 
obtained from initial Gaussian ones as a result of time evolution in the generalized stochastic 
models, where probability distribution functions (pdf's) are obtained from the solutions of the 
differential equations of Fokker-Planck type with so-called fractional derivatives (\cite{gs1,gs2}). 
In other words, the initial Gaussian fluctuations (if any) may become non-Gaussian as a result of 
primordial, fast time evolution. After it, the slower evolution, dictated by the lambda cold dark 
matter ($\Lambda$CDM) scenario, occurs. Although here we do not present the details of this 
primordial time evolution, one of the aims of the present paper is to draw attention to the method, 
which permits to calculate the non-Gaussian distribution function, based solely on the form of 
interaction between astronomical objects. This distribution function is a terminal function for 
above initial fast time evolution process.

In hierarchical clustering type of scenarios, the large scale structure forms from bottom to 
top as a consequence of gravitational interactions between the constituents.
This means that the smaller structures like galaxies are formed first with their later 
merger into larger ones. Consequently, the galaxies spin angular momenta arise as a result of tidal interaction with their neighbours (\cite{Sch09}).  In the originally hierarchical 
clustering  scenarios, the completely random distribution of galaxies angular momenta has been 
obtained. Note however, that it has been shown later that the local tidal shear tensor can generate a local alignment of galaxies angular momenta (\cite{cat96,ca96,Lee02,Nav04}). The mechanisms of this type are known as tidal torque mechanisms, which had first been introduced by \cite{hoyle} and later by \cite{white}, see also the review paper of \cite{kiess15} that discusses galaxies alignments in the context of  gravitational lensing.  Note, that in our model  the angular momentum is the result of tidal interaction with the entire environment, which occurs via interaction transfer from close to distant galaxies, see below. In this sence our approach is the improvement of those considered by \cite{sch, cat96,ca96,Lee02}, where the "mean" tidal interaction with the entire environment has been considered.

The above  tidal torque mechanism has an opposite idea, constructed  on the base of Zeldovich 
pancake model (\cite{Sunyaew72,Doroshkevich73,Shandarin74}). In this model, the  structures in 
the Universe arise from top to bottom. The crucial role here plays a magneto-hydrodynamic shock 
wave which makes the large structure to fragment. This shock wave arise as the result of 
asymmetrical collapse of initial large structure and also imparts galaxies with spin angular momentum. 
The model predicts a coherent, non-random spatial orientation of galaxy planes with the galaxies 
rotational axes to be parallel to the main plane of a large structure.

In the model of primordial turbulences, the spin angular momentum is a remnant of the primordial whirl 
(\cite{vWe51,Gam52,Oze78,Silk83}). As result it is obtained that the rotational axes of galaxies are
oriented not randomly. The preferred direction of angular momentum of galaxies is perpendicular to the 
initial large structure's main plane.

It had been pointed out in (\cite{Gam46,Goe49}) and later in (\cite{Coll73}), that if the Universe is 
rotating, the emerging galaxies angular momentum is a consequence of its conservation in a rotating Universe. 
At that time, the argument against was that this model predicts the galaxies rotational axes alignment, 
which had not been confirmed observationally (see \cite{g2011a} for details). Based on this idea, 
\cite{Li98} introduced  a model in which galaxy forms in a rotating Universe.

We emphasize, that simple picture, where each of the above approaches (primordial turbulences, hierarchical 
clustering and Zeldovich pancakes) predict different ways of galaxies rotational axes ordering is not completely 
true. The point is that in each of the above models including hierarchical clustering, the phase with shock 
wave can appear. Latter phase is usually accompanied by the collapse of structures or substructures 
(\cite{Mel89,Sah94, Pau95,MO05,Shandarin12}), which may generate the rotational axes ordering. Apparently, 
the scale of such orientation is different in different models. For instance, in the Bower's scenario (\cite{Bower05}),
we do not have hierarchical clustering for all scales of masses. Instead, we have anti - hierarchical clustering 
in the small scales as tidal interaction effects yield Zeldovich pancake - like objects emergence (\cite{Zeldovich70}) 
rather then spherically collapsing haloes. There is, however, a fundamental difference with above classical pancake 
scenario. Namely, the anti-hierarchical clustering is local as it occurs in small scale.

The model of hierarchical clustering is the only model explicitly taking into account the dark matter existence. 
The Li model has originally considered the Universe as dust fluid, however, nothing prevents to introduce the dark matter as a background. As a result, in this model, the dark matter is not interacting with observable matter in any 
other way than gravitational forces. In the remaining models, namely primordial turbulences and Zeldovich 
pancake model, the only dust component has been considered so that there are no clear and successful attempts 
to introduce the dark matter there. Therefore, we exclude both models from the present consideration.

Theoretical models of galaxy formation have problems with explaining the observational dependence between 
structure angular momentum and its mass. This dependence can be seen only in two classes of models. There 
are the tidal torque scenario (\cite{heav88,cat96,Hwang07,Noh06a,Noh06b}) and Li model (\cite{Li98,Godlowski05}).
The remaining models do not anticipate such dependence.

Comparing the two models, we should note that Li model needs a global or at least large scale rotation of 
the Universe. \cite{Li98} studied the dependence between the angular momentum and the mass of spiral galaxies 
and he estimated the rotation of the Universe to be close to the value obtained by \cite{Birch82}.
However, the obtained value is too large compared to observed anisotropy in Cosmic Microwave Background Radiation (CMBR). Hence, in the present paper, we consider the Tidal Torque scenario only.

In the present work we perform the comprehensive theoretical analysis of the influence of tidal 
interaction between astronomical objects on the larger (then initial constituents) structures 
formation. The idea of our approach is to use the statistical method originally proposed by 
\cite{chan43}, where we account also for dark matter haloes. The statistical method of 
\cite{chan43} permits to derive the distribution functions of gravitational fields and angular momenta 
of stellar components. Our main result is that in the stellar systems with multipole (tidal) gravitational 
interaction, the derived distribution function cannot be Gaussian. Instead we obtain the pdf which rather 
belongs to the family of so-called "heavy-tailed distributions" (\cite{gs1, gs2, kapkes92}).
As we have mentioned above, the obtained non-Gaussian pdf is a result of fractional time 
evolution for initial Gaussian fluctuations. 
This function allows us to calculate the distribution of virtually any observables (like angular momentum) 
of the astronomical structures (not only galaxy clusters but smooth component like haloes, which mass 
dominate the total mass of the cluster, see \cite{kb12}) in any (linear or nonlinear) Eulerian approach.

The paper is organized as follows. To make the paper self-contained, in the section 2 we shortly recollect
our method (\cite{myapj}) putting more impact to its points, important for present consideration.
Some technical details are described in the appendix. In the section 3 we discuss the 
problem of angular momenta pdf. We show that different (physically reasonable) assumptions about  
the structure of galaxy clusters generate different relations between their mass $M$
and average angular momentum $L$. We show that it is possible to derive not only the relation 
$L \sim M^{4/3}$ (like in  \cite{myapj}) but also recover well-known 
empirical relation $L \sim M^{5/3}$. We show that while 
it is possible to discriminate between the above model assumptions theoretically, the present observational data 
are not sufficient to come to unambiguous conclusion. 
We also discuss the possibility of observational testing of our 
theoretical results related to the time evolution of  the distribution function of angular momenta and 
its mean value $L$. We conclude our article by the section 4.

\section{Distribution function of gravitational fields}

We consider the tidal interaction in the ensemble of galaxies and their clusters in the 
Friedmann - Lema\^{\i}tre - ?Robertson - ?Walker Universe with Newtonian self-gravitating dust fluid ($p=0$) 
containing  both luminous and dark matter. 
The tidal (shape distorting) interaction between the astronomical objects can be derived by the multipole 
expansion of the Newtonian interaction potential between fluid elements (\cite{poisson98}). Limiting ourselves 
to quadrupolar term, we write the Hamiltonian function of interaction between above elements in the form 
\begin{equation}
{\cal H}=- G\sum_{ij}Q_im_jV({\bf r}_{ij}),\ 
V({\bf r})=\frac 12 \frac{3\cos^2\theta - 1}{r^3}, \label{star1}
\end{equation}
where $G$ is the gravitational constant, $Q_i$ and $m_i$ are, respectively, the quadrupole moment and mass of $i$-th 
object, $r_{ij}\equiv $ $|{\bf r}_{ij}|$, ${\bf r}_{ij}=$  ${\bf r}_{j}-$ ${\bf r}_{i}$ is a relative distance 
between objects while $\theta$ is the apex angle. Our Hamiltonian function \eqref{star1} is obtained  for the ensemble of $N$ objects, thus generalizing the result of  \cite{poisson98} for two particles. 

Note, that the Hamiltonian function \eqref{star1} describes the interaction of quadrupoles, formed both from luminous and dark matter. This is important as in real world the galaxies, formed from luminous matter, reside inside dark matter halos that are much more extended and massive. In other words, the Hamiltonian function \eqref{star1} (and subsequent results) already contains the information about dark matter haloes. We have discussed this question in our previous work \cite{myapj}. The main point was that the properties of luminous matter (like galaxies and their clusters) give us information about those of dark matter (sub)structures. This point is corroborated by observations (see, e.g. \cite{Paz08,Bett10,Kim11,Varela12}) that angular momentum of luminous matter is correlated with that of corresponding dark matter haloes. Below we will calculate the angular momentum of luminous astronomical structures.
Our formalism can be generalized to describe not only this situation, but the structures with larger smooth component. Namely, in general, the luminous galaxy matter is not only surrounded by dark matter haloes, but also (along with latter haloes) submerged in the "mud", which is hypothetical intergalaxies dark matter. We plan to fulfill this interesting generalization in our subsequent publications. 

In the function \eqref{star1}, we split the interaction between many stellar objects (particles) to that in pairs, see Appendix A for details. Such splitting is usual, 
for instance in the theory of magnetism, where the interacting spins ensemble is represented by the sum of all possible 
couplings between particle pairs $i$ and $j$. For instance, the three particle interaction may be decomposed as 123 = 12 + 13 + 23, see, e.g.  \cite{mat}.

The Hamiltonian function \eqref{star1} describes the pairwise, shape-distorting interaction between the 
structures. Namely, this interaction distorts  the shape of a given $i$-th object, which alters its density field 
$\rho_i({\bf r})$. As the objects have random shapes, their masses $m_i$ and quadrupole moments $Q_i$ vary randomly 
likewise the gravity field ${\bf E}_{quad}$ from these quadrupoles. One should note that latter field is in fact a 
gradient of the potential, given by equation \eqref{star1}.  It has the form:
 \begin{equation}\label{star3}
 {\bf E}_{quad}({\bf r})={\bf i}_rE_0\frac{3\cos^2\theta - 1}{r^4},
 \end{equation}
where $E_0=GQ/2$ and ${\bf i}_r$ is the unit vector in radial direction.
 
According to statistical method of \cite{chan43}, the distribution function of
random quadrupolar fields is  
\begin{equation}\label{star8}
f({\bf E})=\overline{\delta({\bf E}-{\bf E}_i)},
\end{equation}
where $\delta(x)$ is Dirac $\delta$ - function, while  ${\bf E}_i\equiv$ ${\bf E}_{quad}({\bf r}_i)$ is 
given by Eq. \eqref{star3} where the bar means averaging over spatial (and any other possible) disorder. 
Moreover, if all objects in the ensemble are similar (no randomness), then the distribution function is represented 
by the simple delta-peak, centered at the field ${\bf E}_i$. The disorder broadens this delta-peak, giving rise to 
"bell-shaped" continuous probability distribution, see \cite{stef97,semstef02,semstef03} and references therein.
 
The explicit averaging in Eq. \eqref{star8} is performed with the help of the integral representation of Dirac $\delta$ 
- function, see  \cite{myapj} for details. The idea is that the mass and quadrupole moment of the object in the 
volume $V$ obey the uniform distribution with probability density equal to $1/V$. In such a case we introduce the 
number of objects $N$ so that in the limit $N \to \infty$ and $V \to \infty$, their density $n=N/V$ remains constant. 
Final expression for the distribution function \eqref{star8} reads (\cite{myapj})
\begin{eqnarray}
&& f({\bf E})=\frac{1}{(2\pi)^3}\int_{-\infty}^\infty\ e^{i{\bf E}{\boldsymbol \rho}-F(\rho)}d^3\rho,\label{star12} \\
&& F(\rho)=
n\int_V\left[1-\frac{\sin \rho E({\bf r})}{\rho E({\bf r})}\right]d^3r. \label{star12f} 
\end{eqnarray}
In this case $F(\rho)$ is in fact the characteristic function for random gravitational fields distribution. 
Note also that characteristic function $F(\rho)$ depends only on modulus $\rho$ and not its angles. This will result 
(see Eq. \eqref{star19} below) in the only field modulus dependence of pdf of random gravitational fields. The reason is 
that we take only $zz$ component of quadrupolar field in Eq.\eqref{star3}. If we need the complete (i.e. including its 
possible angular dependence) distribution function of vector ${\bf E}$, we should account for complete tensor structure of Hamiltonian \eqref{star1} ${\cal H}=- G\sum_{ij\alpha\beta}Q_{i\alpha\beta}m_jV_{\alpha \beta}({\bf r}_{ij})$, $\alpha,\beta=x,y,z$. 
Such account (\cite{stef97,semstef02,semstef03}), while not changing our conclusions qualitatively (and in many cases 
quantitatively, see below), will make the problem to be tractable only numerically. At the same time our present approach permits to gain analytical insights into the problem (for example investigate the implication of non-Gaussian character of distribution function of gravitational fields), which is good starting point for future numerical simulations. One more justification of the radial distribution is the results of numerical simulations in halo model (\cite{sb10}), where the axes of galaxies embedded in dark matter halo, were preferentially radially oriented.

Moreover, the spin angular momentum is usually known only for very few galaxies and other structures. For this reason,  the spatial orientation of galaxies (see, for example, \cite{f4,Rom12}) is studied instead of their angular momenta. Alternatively, only the distribution of position angles of galaxy planes is analysed in \cite{h4}.

In more realistic models of galaxy clustering we can assume that the stellar objects like galaxies density is 
not a constant but rather depends on their separation $n=n({\bf r})$. The other factor, which may improve the coincidence with observational results is to consider the galaxy clustering within a model of inhomogeneous distribution of masses (and/or quadrupolar moments) in the large scale structure. The idea here is to introduce the distribution function of masses $\tau(m)$, which had been put forward by \cite{chan43}.

It is important that distribution function $f({\bf E})$ \eqref{star12} in general case could be much more complicated than 
simple Gaussian. We had shown in  \cite{myapj}  that for multipole interaction between astronomical objects, the function \eqref{star12} does not admit Gaussian limit. The calculation of $F(\rho)$ \eqref{star12f} generates following explicit form of $f(E)$ (\cite{myapj})
\begin{eqnarray}
&&f(E)=\frac{1}{2\pi^2E}\int_0^\infty \rho \ e^{-\alpha \rho^{3/4}}\ \sin \rho E\
d\rho,\label{star19}\\
&& \alpha= 2\pi n \cdot 0.41807255\cdot E_0^{3/4}. \nonumber
\end{eqnarray}
The expression \eqref{star19} is the chief theoretical result of our studies.  The distribution function 
\eqref{star19} depends parametrically on the objects (i.e. both luminous and dark matter) density $n$,  and on 
average quadrupole moment $Q$. 

The normalization condition for distribution function \eqref{star19} reads
\begin{equation}\label{sta1}
4\pi \int_0^\infty E^2f(E)dE=1.
\end{equation}

As we have shown previously (\cite{myapj}), the distribution function of the gravitational fields cannot be Gaussian 
for multipole interaction between galaxies or any other astronomical objects including elements of dark 
matter halos. However, all previous theories postulated the distribution function in the Gaussian form rather then 
calculated it. We mention here that non-Gaussian distribution have also been postulated rather then calculated in \cite{bart04}. 
In our opinion, non-Gausssian, heavy-tailed nature of the above pdf captures the essential physics 
of the systems with long-range gravitational multipole interaction. Namely, the long-range interaction in such 
systems makes the objects (galaxies, their clusters and even the dark matter halos) to interact with each other 
also at very large separations. This, in turn, implies nonzero probabilities of such configurations, contrary to 
the case of Gaussian distribution, generated by short-range interactions. Below we will see the important implications 
of this fact.

To plot the function $f(E)$, we define the dimensionless variables  
$\rho E=x$ and $\beta= E/\alpha^{4/3}$.In these variables the integral \eqref{star19} assumes the form:
\begin{eqnarray}
f(\beta)=\frac{H(\beta)}{4\pi\beta^2\alpha^4},\
H(\beta)=\frac{2I(\beta)}{\pi \beta}, \label{star20a}\\
I(\beta)=\int_0^\infty x \sin x \exp{\left[-\left(\frac x\beta\right)^{3/4}\right]}dx. \label{star20b}
\end{eqnarray}
The physical interpretation  of the function $H(\beta)$ is following. This function gives the effective 1D distribution 
function of random gravitational fields. This is because the normalization condition for $H(\beta)$ is of effectively one-dimensional form $\int_0^\infty H(\beta)d\beta=1$, see \eqref{sta1}. In this case, the average value  ${\bar \beta}$ of dimensionless random field $\beta$ has the form ${\bar \beta}=\int_0^\infty \beta H(\beta)d\beta$. The mean value 
${\bar \beta}$ exists if the integral $H(\beta)$ is convergent. The asymptotic analysis of the function $f(\beta)$, which had been performed in our previous work (\cite{myapj}), shows that $f(\beta)$ does not depend 
on $\beta$ for small $\beta$ and decays  as $\beta^{-7/4}$ at large $\beta$. The character of decay at large $\beta$ shows that although normalization integral is convergent, already first moment does not exist. Such behavior is a characteristic feature of so-called heavy-tailed distributions (\cite{gs1,gs2}).

\section{Distribution function of angular momenta}

Our aim is to derive the distribution function of angular momenta. For that we need to calculate  how the angular 
momentum ${\bf L}$ of a stellar object depends on its gravitational field ${\bf E}_{quad}({\bf r})$ \eqref{star3}. The 
expression for angular momentum components $L_\alpha$ ($\alpha=x,y,z$) could be obtained perturbatively in small Lagrangian coordinate ${\bf q}$. One should note that the first order terms were obtained in Eq. (11) of  \cite{cat96}, while the second order ones in their next article \cite{ca96} (Eq.  (28)). Note that both equations have identical structure i.e. 
$L_\alpha^{(i)}=f_i(t)\varepsilon_{\alpha \beta \gamma}E_{i \beta \sigma}I_{\sigma \gamma},\ \alpha,\beta,\gamma,\sigma = x,y,z$, where index $i=1,2$ defines the order of perturbation theory, $\varepsilon_{\alpha \beta \gamma}$ is Levi-Civita symbol, $E_{\beta \sigma}$ are components of quadrupole (tidal) field \eqref{star3} while $I_{\sigma \gamma}$ represent the components of inertia tensor.

In order to obtain the distribution function of {\em modulus} of $E$ (and subsequently $L$), it is sufficient to take $zz$ 
component in \eqref{star3}. If we need the complete distribution function of vector ${\bf E}$, we should account for complete tensor structure of Hamiltonian \eqref{star1} 
${\cal H}=- G\sum_{ij\alpha\beta}Q_{i\alpha\beta}m_jV_{\alpha \beta}({\bf r}_{ij})$, $\alpha,\beta=x,y,z$. Also, as 
${\bf L}$ is a function of time $t$ by means of the functions $f_i(t)$, the distribution function will be time dependent.
With respect to symmetry relations $I_{ab}=I_{ba}$ and $E_{ab}=E_{ba}$ and leaving only $E_{zz}$, we obtain $L_x=-b(t)E_{zz}I_{yz}$, $L_y=b(t)E_{zz}I_{xz}$, $L_z=0$, $L=\sqrt{L_x^2+L_y^2+L_z^2}= L_0E$, $L_0=L_0(t)=f_i(t)\sqrt{I_{xz}^2+I_{yz}^2}$. Above equations constitute linear relation between angular momentum and tidal field moduli. They are valid both in linear ($i=$1) and nonlinear ($i=$2) regimes. Because above relation between gravitational 
field modulus and angular momentum is linear in both cases, it is easy to see that the shape of distribution function of angular momenta $f(L)$ repeats that of gravitational fields. In the explicit form  expression for $f(L)$ can be derived using well known relation from the theory of probability $f(L)=f[E(L)]\left|\frac{dE}{dL}\right|$, which yields
\begin{equation}\label{star26}
f(L)=\frac{1}{2\pi^2 L}\int_0^\infty \rho \ e^{-\alpha \rho^{3/4}}\ \sin \left(\rho\frac{L}{L_0(t)}\right) \ d\rho,
\end{equation}
where $L_0(t)$ is defined above. Dimensionless variables $\rho (L/L_0)=x$, $\lambda=L/(L_0\alpha^{4/3})$ generate 
the pair of functions which are similar to those obtained for the gravitational fields distribution. They read:
\begin{equation}\label{star27a}
f(\lambda)=\frac{H(\lambda)}{4\pi\lambda^2\alpha^4L_0},\ H(\lambda)=\frac{2I(\lambda)}{\pi \lambda},
\end{equation}
where $I(\lambda)$ is defined by the expression \eqref{star20b} {{and is usually referred to as spin parameter.}}

\begin{figure}
\centering
\includegraphics[width=0.8\columnwidth]{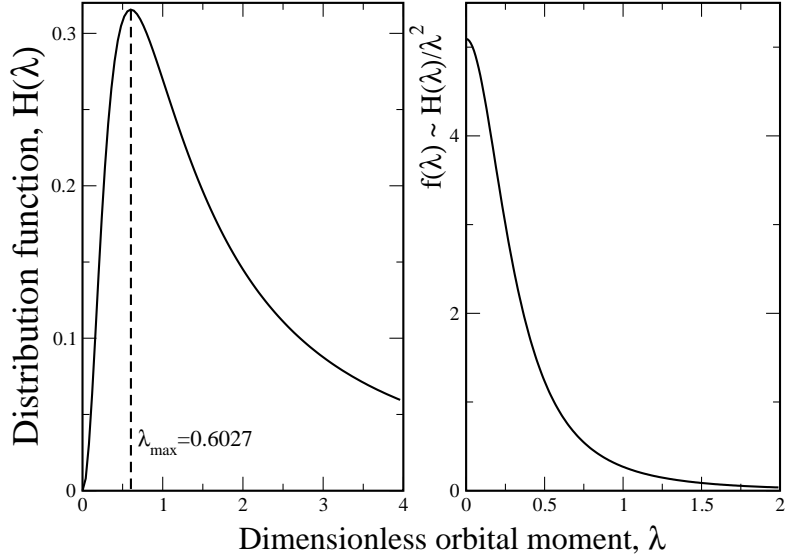}
\caption{Left panel shows the effective 1D distribution function  $H(\lambda)$ \eqref{star27a}. The shape of the function is the same as the distribution function \eqref{star20b}. Dashed line represents the  value of argument 
$\lambda_{max}$, related to maximum of $H(\lambda)$. Right panel shows 3D distribution function $4\pi \alpha^4 L_0f(\lambda)=$ $H(\lambda)/\lambda^2$.} \label{fig1}
\end{figure}

The effective 1D distribution function for gravitational  fields or angular momenta is presented in left panel of Fig. \ref{fig1}. 
It is seen that while initial 3D function $f(\lambda)$ decays monotonically (right panel), this function is strongly asymmetric 
and has characteristic bell shape. Note, that as the initial equation \eqref{star1} allows for the interaction between all astronomical objects in an ensemble, it considers naturally the interaction with surrounding structures and dark matter haloes also. This fact renders the distribution functions of gravitational fields \eqref{star20a} and angular momenta 
\eqref{star27a} to account not only for the isolated cluster regions, but for long-range interactions with surrounding structures as well. To be specific, the narrow peak of distribution function in left panel of Fig.\ref{fig1}  stems from the closely situated cluster region, while its long tails stem from the long-range (quadrupole) interaction with surrounding structures. In other words, the interaction with surrounding structures is essential (and our distribution functions takes this fact into account) as the interaction between objects in stellar ensembles have long-range multipole character.

As we have shown in the previous article (\cite{myapj}), the integral for the first moment of angular momenta pdf diverges. It is well - known (see, for example, \cite{kapkes92}) 
that for the distribution functions, which decay slowly at infinities, the corresponding mean value can be approximately estimated as the maximum of such function. In this spirit we calculate $\lambda_{max}$, corresponding to the maximum of distribution function $H(\lambda)$,  as presented on the Fig. \ref{fig1}. The analysis of $\lambda_{max}$ in dimensional units makes possible to obtain 
some useful relations, which earlier had been guessed only empirically. To consider the characteristics of galaxies, i.e. 
luminous matter, here we use the ideas of halo model (\cite{sb10}), which states that galaxies (i.e. "pieces" of luminous matter) are embedded in the dark matter haloes so that their observable characteristics like angular momentum emerge from the mass and hence gravitational field of dark matter. Also, as the galaxies and their clusters reside in the larger structures like voids and filaments, the gravitational field of latter large objects also influence galaxies, see, e.g.  \cite{Joachimi15}. 
As our distribution function \eqref{star27a} takes these effects into account by virtue of model \eqref{star1}, our subsequent 
calculations of mean angular momentum of the galaxies take above effects into account.

Let us first consider the simplest possible CDM model in the first order of perturbation theory. In such model 
the evolution of scale factor is given by the equation $a(t)=D(t)=(t/t_0)^{2/3}$ (\cite{dor70}) 
so that $L_0=\frac{2I}{3t_0}\tau$, $\tau=t/t_0$, $I=\sqrt{I_{xz}^2+I_{yz}^2}$. 
The equation $dH/d\lambda=0$ has solution $\lambda_{max}=0.602730263$, which give in dimensional units
\begin{eqnarray}
&&L_{max}=0.7281884n^{4/3}\frac{t}{t_0^2}GIQ \approx \nonumber \\
&&\approx \kappa n^{4/3}\frac{t}{t_0^2}GR^4m^2, \label{lm2}
\end{eqnarray}
where $n=N/V$, $\kappa \sim 1$ is a constant. To derive the equation \eqref{lm2}, we estimate galaxy 
quadrupole moment $Q$ and its mean inertia moment $I$ as being proportional to $mR^2$, where $m$ is mass of galaxy while $R$ is its mean radius. In our approach we represent volume $V$ as $V=R^3$, then $R$ cancels down in Eq. \eqref{lm2} so that $L_{max} \sim (t/t_0^2) m^2 N^{4/3}$. Then, we introduce the mass of a galaxy cluster $M=mN$ and obtain 
 \begin{equation}\label{lm2a}
L_{max} \sim \frac{t}{t_0^2} M^{5/3} \left(\frac mN\right)^{1/3}\equiv \frac{t}{t_0^2} M^{5/3} \frac{\rho^{1/3}}{n^{1/3}},
\end{equation}
where $\rho=m/V$ is a mass density and $n=N/V$ is galaxies density. 
Following \cite{cat96}, we assume that mass density $\rho$ is a function of time, defined by Friedmann equation in CDM model 
${\dot a}/a=H_0 =$ $\sqrt{8\pi G\rho/3}$, where $H_0$ is the Hubble constant. 
This generates the dependence $\rho \propto$ $t^{-2}$, which, being substituted to 
Eq. \eqref{lm2a}, yields
\begin{equation}\label{lm2b}
L_{max} \sim \frac{t^{1/3}}{t_0^2} \frac{M^{5/3}}{n^{1/3}} \sim t^{1/3} M^{5/3}.
\end{equation}
To derive Eq. \eqref{lm2b}, we assume that $n=$ const. We see that equations \eqref{lm2a} and \eqref{lm2b} recover the expression (27) of  \cite{cat96}, giving the theoretical derivation of well-known empirical relation between mean value angular momentum of galaxies ensemble (galaxy clusters) and their mass $L_{max} \sim M^{5/3}$ (see \cite{cat96} and references therein). {{Note, that within tidal torque model the $M^{5/3}$ - law has been first obtained by \cite{heav88} while reasonable values for lambda in Eq. \eqref{star27a} within the tidal torque approach has been derived by \cite{sch}, who followed  \cite{heav88}.}}

There is also other approach to interpret the dependence of $L_{max}$ on stellar parameters. Namely, suppose that volume $V=R_A^3$, where $R_A$ is a mean cluster radius,  proportional to the autocorrelation radius (see \cite{long08} and references therein). In such  approach (see \cite{myapj} for details)  $n$  is still a constant for any particular cluster, but now it varies from cluster to cluster with increasing richness $N$. In this case we may rewrite $N=M/m$ to obtain the 
alternative (to Eq. \eqref{lm2b}) form of expression for $L_{max}$
\begin{equation}\label{lm3}
L_{max} \sim \frac{t}{t_0^2}\ \left(\frac{R}{R_A}\right)^4\ m^{2/3}M^{4/3},
\end{equation}
which does not contain $\rho$. 

It is instructive to comment on time dependence $L_{max}(t)$ in Eq. \eqref{lm3}. On the first sight, 
it follows from \eqref{lm3} that $L_{max} \sim t$, but the problem complicates a lot by the intricate
time dependence of the quantities $R$ and $R_A$ (\cite{long08}). We plan to study this question in future works. 

It is clear from the  equation \eqref{lm2} that mean orbital moment of a galaxy increases with the number of galaxies $N$ and it is proportional to $N^{4/3}$. Moreover, even in the model with  constant galaxies density $n$, number (richness) $N$ varies from cluster to cluster so that the dependence $L_{max}(N)=$ $\kappa_2N^{4/3}$ holds and shows that angular momenta increase with number of galaxies $N$ in analysed structure.

The sample of 247 Abell cluster has been analysed by \cite{g10a}. Namely, the orientation of galaxies in 
particular clusters has been studied. The idea was to test hypotheses that the galaxies angular momenta increase with the cluster richness. If galaxy cluster do not rotate (see \cite{regos98, Hwang07}), then increasing alignment of galaxies in clusters mean the increase of the angular momentum of whole cluster. In the paper of \cite{g10a} the orientation of galaxies was quantified by distribution of the angles. Specifically, the position angle of galaxy plane $p$ and two angles $\delta_d$, giving spatial orientation of the normal to galaxy plane, have been considered. The authors have also studied two additional angles. One is the angle between the normal to the galaxy plane and the main plane of the coordinate system. The second is the angle $\eta$ between the projection of this normal onto the main plane and the direction toward the zero initial meridian (\cite{f4}). 

The entire range of all investigated angles was arranged into $n$ bins. As we would like to detect non-random efect in the galaxies orientation, we first check whether the orientation is isotropic. To be specific, we check if the disribution of analyzed angles in the clusters under investigation is isotropic. 
The distribution of the above angles has been investigated using the statistical tests. 
They were $\chi^2$ and the Fourier Test. However, in the present paper we 
extend the analysis for first auto-correlation and Kolmogorov-Smirnov tests (K-S test) (\cite{h4,f4,g10a,g2011b}). 

The statistics $\chi^2$ is:
\begin{equation}
\label{eq:c1}
\chi^2 = \sum_{k = 1}^n {(N_k -N\,p_k)^2 \over N\,p_k}= \sum_{k = 1}^n {(N_k -N_{0,k})^2 \over N_{0,k}},
\end{equation}
where $p_k$ are probabilities that chosen galaxy falls into $k$-th bin, $N$ is the total number 
of galaxies in a sample (in our case in a cluster), $N_k$ is the number of galaxies within the $k$-th  
angular bin and $N_{0,k}$ is the expected number of galaxies in the $k$-th bin. 

The first auto-correlation test quantifies the correlations between galaxy numbers in neighboring angle bins. 
The statistics $C$ reads
 
\begin{equation}
\label{eq:c2}
C\, = \, \sum_{k = 1}^n { (N_k -N_{0,k})(N_{k+1} -N_{0,k+1} )
\ \over \left[ N_{0,k} N_{0,k+1}\right]^{1/2} },
\end{equation}
where $N_{n+1}=N_1$.

If the deviation from isotropy is a slowly varying function of the analyzed angle $\theta$, 
one can use the Fourier test:

\begin{equation}
\label{eq:f9}
N_k = N_{0,k} (1+\Delta_{11} \cos{2 \theta_k} +\Delta_{21} \sin{2
\theta_k}+\Delta_{12} \cos{4 \theta_k}+\Delta_{22} \sin{4\theta_k}+.....).
\end{equation}
In this test, the crucial statistical quantities are amplitudes
\begin{equation}
\label{eq:f6}
\Delta_1 = \left( \Delta_{11}^2 + \Delta_{21}^2 \right)^{1/2},
\end{equation}
(only the first Fourier mode is taken into account) or 
\begin{equation}
\label{eq:f16}
\Delta =
 \left( \Delta_{11}^2 + \Delta_{21}^2+\Delta_{12}^2 + \Delta_{22}^2 \right)^{1/2},
\end{equation}
where the first and second Fourier modes are analysed together.
During our investigations we analyzed statistics $\Delta_1/\sigma(\Delta_1)$
and $\Delta/\sigma(\Delta)$ (see \cite{g10a} for details).

In the case of K-S test we investigate statistics $\lambda$:
\begin{equation}
\label{eq:k1}
\lambda=\sqrt{N}\,D_n
\end{equation}
which is given by limiting Kolmogorov distribution, where
\begin{equation}
\label{eq:k2}
D_n= \sup|F(x)-S(x)|
\end{equation}
and $F(x)$ and $S(x)$ are theoretical and observational distributions of the investigated angle respectively.

The aim of the paper of \cite{g10a} was to test the hypotheses that alignment of galaxies increases with cluster 
richness (\cite{Godlowski05}). The main result of \cite{g10a} was that the values of investigated statistics
increase with increasing number of cluster galaxy members. This permits to conclude that there exist a relation between anisotropy and the number of galaxies in a cluster. Note, that above testing has been performed assuming linear model $y = aN+b$ where $y$ is a value of investigated statistics, $N$ is the cluster members number while $a$ and $b$ are linear regression coefficients. In the paper of \cite{g10a}, the null hypothesis $H_0$ (that the investigated statistics is random one, i.e. neither increases nor decreases so that parameter $t=a/\sigma(a)=0$) has been confronted against $H_1$ hypothesis that statistics increases with the cluster richness i.e. $t>0$. 
In our previous paper (\cite{myapj}), as well as in the present paper, we show that dependence 
between the alignment of galaxies in clusters and number of members galaxies is not necessarily linear but could be, according to above assumptions as either $N^{4/3}$ or $N^{5/3}$.

\begin{figure}
\begin{center}
\includegraphics[width=0.9\columnwidth]{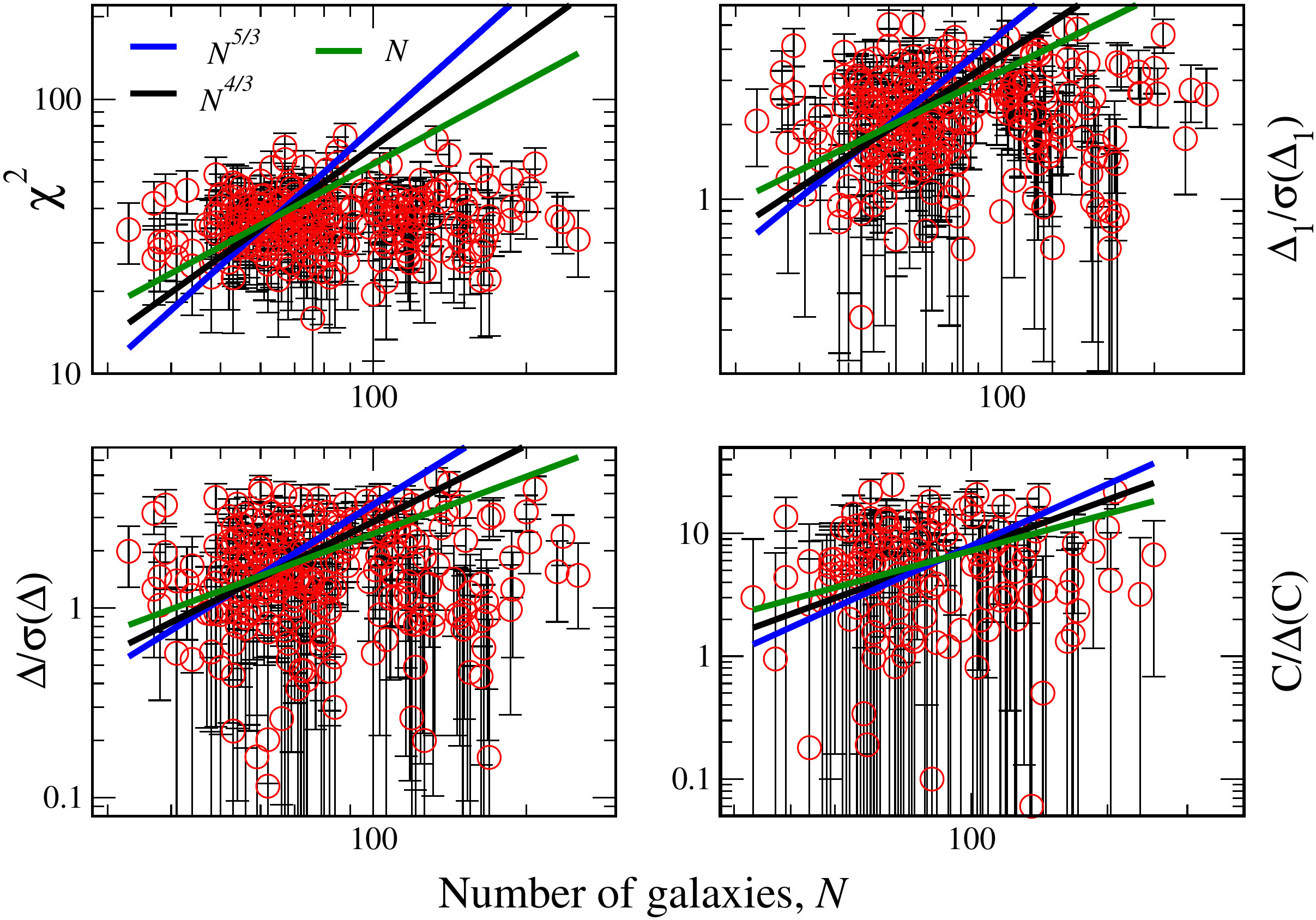}
\end{center}
\caption{The dependence between the number of galaxies in the cluster $N$
and the value of analyzed statistics ($\chi^2$ - left upper panel, $\Delta_1/\sigma(\Delta_1)$ - right upper panel, $\Delta/\sigma(\Delta)$ - left lower panel, $C/\sigma(C)$ - right lower panel) for the position angles $p$. Mind double log scale, chosen to make the dependences $N^\alpha$ ($\alpha=$1, 4/3, 5/3) straight lines.}  
\label{gala}
\end{figure}

For this reason, in the figure \ref{gala}, we present statistics
($\chi^2$, Fourier and First autocorrelation tests \cite{h4,f4,g10a}) for the case of the position angles 
obtained for the sample of $247$ rich Abell clusters, analysed by \cite{g10a}. We present  linear dependence $\sim N$ as well as the cases when analysed statistics increases as $N^{4/3}$ and $N^{5/3}$. 
The error bars presented in the figure \ref{gala}, suggest that the data points seem
to be not sufficient do discriminate between models. For this reason we analyze the dependence 
between the number of galaxies in a cluster and the value of analyzed statistics in more details.
We performed the investigation of the linear regression given by  $y=aN+b$
counted for various parameters.  Namely, we have studied  the linear regression between
different statistics $\chi^2$, $\Delta_1/\sigma(\Delta_1)$, $\Delta/\sigma(\Delta)$, $C$ or $\lambda$
and the number of analyzed galaxies in each particular cluster. This has been done for the case of linear 
dependence $\sim N$ or power laws $\sim N^{4/3}$ and $\sim N^{5/3}$ in the case of remaining models.

Now we  assume that the theoretical, uniform, random distribution contains the same number 
of clusters as the observed one. To be specific, we consider null hypothesis $H_0$ that
the distribution is a random one and neither increases nor decreases. This means that expected value 
of statistics $t=a/\sigma(a)=0$, while $t$ statistics has Student's  distribution with $u-2$ degrees 
of freedom, where $u$ is the number of analyzed clusters. In other words, we test $H_0$
hypothesis that $t=0$ against $H_1$ hypothesis that $t>0$. Of course, in order to reject
the $H_0$ hypothesis, the value of the observed statistics $t$ should be
greater than $t_{cr}$ which we could obtain from staistical tables. Note that for our
sample containing only 247 clusters, the critical value at the significance level $\alpha=0.05$ is equal to
$t_{cr} = 1.651$.

The result of our statistical analysis is presented in the Table \ref{tab:t1}. 
We analysed two samples of data. In the first sample $A$ all galaxies lying in the area 
regarded as a cluster, were taken into account. In the second sample $B$, to
avoid the "contamination" by the background objects, we restrict ourselves by consideration of the 
galaxies brighter than  $m_3+3$ only.

Note, that the cases of linear dependence for statistics of $\chi^2$, $\Delta_1/\sigma(\Delta_1)$
and $\Delta/\sigma(\Delta)$ have usually been analysed in the paper of \cite{g10a} (Table 1). 
Note, however, that our present results are somewhat different from those obtained by \cite{g10a}. 
For example in the case of $\chi^2$ 
instead of $t=0.025/0.015=1.67$ we obtain $t=1.87$. The reason is that in the paper of \cite{g10a}
the error bars of the data points (i.e. statistics for individual clusters) has been estimated from the 
sample, while now it is taken from exact theoretical analysis (\cite{g2011b,Wang03}).

In majority of cases, exept for the first autocorrelation test,  the values of the
obatined statistics are greater than critical one $t_{cr} = 1.651$. One could observe that 
for all three analyzed models (i.e. linear dependence
$\sim N$ and the increased statistics like $\sim N^{4/3}$ or $\sim N^{5/3}$) we can eliminate
$H_0$ hypothesis (that statistics $t=a/\sigma(a)=0$) in favour of hypothesis $H_1$
that $t>0$. The effect increases if we analyse Sample B which mean that we restrict 
the cluster membership to galaxies brighter than $m_3+3$. The significance of the
effect decreases with increasing powers $m$ in the models like $N^m$, but in 
majority of cases the effect is significant. The above results allow us to conclude that
the presented data is not sufficient to discriminate between above three models so that we need future 
investigations based on the larger cluster samples.

\begin{table}
\begin{center}
\caption{The  statistics $t=a/\sigma(a)$ for 247 rich Abell clusters.
Sample A - all galaxies, Sample B - galaxies brighter than $m_3 + 3$ 
}\label{tab:t1}
\begin{tabular}{cccc}
\hline \hline
Test&$N$&$N^{4/3}$&$N^{5/3}$\\ 
\hline
$Sample A$                       &       &       &        \\
$\chi^2$                         &$1.872$&$1.766$&$1.667$\\
$\Delta_{1}/\sigma(\Delta_{1})$  &$1.613$&$1.588$&$1.580$\\
$\Delta/\sigma(\Delta)$          &$1.964$&$1.941$&$1.821$\\
$C$                              &$1.352$&$1.381$&$1.417$\\
$\lambda$                        &$2.366$&$2.500$&$2.400$\\ 
\hline
$Sample B$                       &       &        &        \\
$\chi^2$                         &$1,979$&$1.801$&$1.625$\\
$\Delta_{1}/\sigma(\Delta_{1})$  &$2.182$&$1.962$&$1.702$\\
$\Delta/\sigma(\Delta)$          &$2.104$&$1.885$&$1.596$\\
$C$                              &$1.225$&$1.170$&$1.125$\\
$\lambda$                        &$2.421$&$2.000$&$1.765$\\
\hline
\end{tabular}

\end{center}
\end{table}

In our investigations, we have also studied the time dependence of galaxies gravitational fields pdf (\cite{myapj}).
The distribution function \eqref{star27a} evolves in time. It relies on explicit 
dependences $f_1(t)$ and $f_2(t)$. The functions $f_1(t)=a^2(t)\dot D(t)$ while $f_2(t)=\dot E(t)$ 
(we use standard notations where dot means time derivative) could be obtained from the differential equations set, derived in $i$ - th order of perturbation theory by \cite{bouchet92}: 
\begin{eqnarray}
&&t_0^2\ddot D(t)+a(t)D(t)=0, \label{first} \\
&&t_0^2\ddot E(t)+a(t)E(t)=-a(t)D(t)^2, \label{sec}
\end{eqnarray}
where $0\leq t <\infty$ is dimensional physical time. The dimensionless function (scale factor) $a(t)$ is determined from the first Friedmann equation. In our investigations we consider the $\Lambda$CDM model, however we compare its predictions with those obtained in classical CDM model.

To obtain the dependence $L_0(t)$, we use substitution $\lambda \to$ $\lambda/f_i(\tau)$,
($\tau=t/t_0$) which yields from \eqref{star27a}
\begin{equation}\label{lcdm1}
H(\lambda,\tau)=\frac{2I(\lambda/f_i(\tau))}{\pi \lambda}, \ i=1,2.
\end{equation}
To derive $f_{1,2}(\tau)$ in particular model ($\Lambda$CDM model in our case), it is necessary to calculate $a(t)$ from the first Friedmann equation, see \cite{myapj} for details:
\begin{equation}\label{frid2}
\frac{da}{dt}=H_0\sqrt{\Omega_\Lambda a^2+\frac{1-\Omega_\Lambda}{a}}.
\end{equation}
The solution of the equation \eqref{frid2} has the form
\begin{eqnarray}
&&a(t)=\alpha \sinh^{2/3}(t/t_0),\ \alpha=\left(\frac{1-\Omega_\Lambda}{\Omega_\Lambda}\right)^{1/3},\label{frid3} \\
&&t_0=\frac{2}{3H_0\sqrt{\Omega_\Lambda}},\nonumber
\end{eqnarray}
where $\Omega_\Lambda=\Lambda/(3H_0^2)$ is cosmological constant or so-called vacuum density, $\Lambda$ is cosmological constant and $H_0$ is Hubble constant.

Having the function $a(t)$, we can solve equation \eqref{first} numerically for $D(\tau)$ and then determine the function
 $f_1(\tau)=$ $a^2(\tau)D'(\tau)$ ($D'=dD/d\tau$). Accordingly, in the nonlinear regime, the function $f_2(\tau)=E'(\tau)$ 
could be calculated numerically from the equation \eqref{sec}.

One should note that functions $f_2(\tau)$, which are related to the second perturbative corrections, are negative. 
For instance, in Einstein - de Sitter model $f_1(\tau)=(2/3)\tau$ and $f_2(\tau)=(-4/7)\tau^{1/3}<0$ (\cite{dor70,ca96}). 
The same result ($f_2(\tau)<0$) can be obtained numerically for $\Lambda$CDM model.

The dependences $H(\lambda,\tau)$ \eqref{lcdm1} for CDM (with above analytical expressions for $f_i(\tau)$) and $\Lambda$CDM models are shown in the Fig. \ref{fig2}. It is easy to observe that as time increases, the distribution function decreases, while its peak grows to infinity at $t \to 0$. As time grows, the whole distribution function "blurs" as its maximum shifts towards large $t$. It is also easy to notice that "blurring" of distribution function at large times is much faster for $\Lambda$CDM model. Also, both in linear and nonlinear regimes $H(\lambda,\tau)$ increases with time. We emphasize once more that in 
$\Lambda$CDM model this growth is much faster than in the CDM model. It is the consequence of the fact that functions $f_i(\tau)$ enter the exponent in the integrand \eqref{lcdm1}. The comparison of upper and lower panels of Fig. \ref{fig2} show that the behaviour of $H(\lambda,\tau)$ is qualitatively similar in linear and nonlinear regimes of fluctuation growth. This leads to conclusion that even linear regime give qualitatively correct approximation to the function $H(\lambda,\tau)$.

\begin{figure}
\centering
\includegraphics[width=0.8\columnwidth]{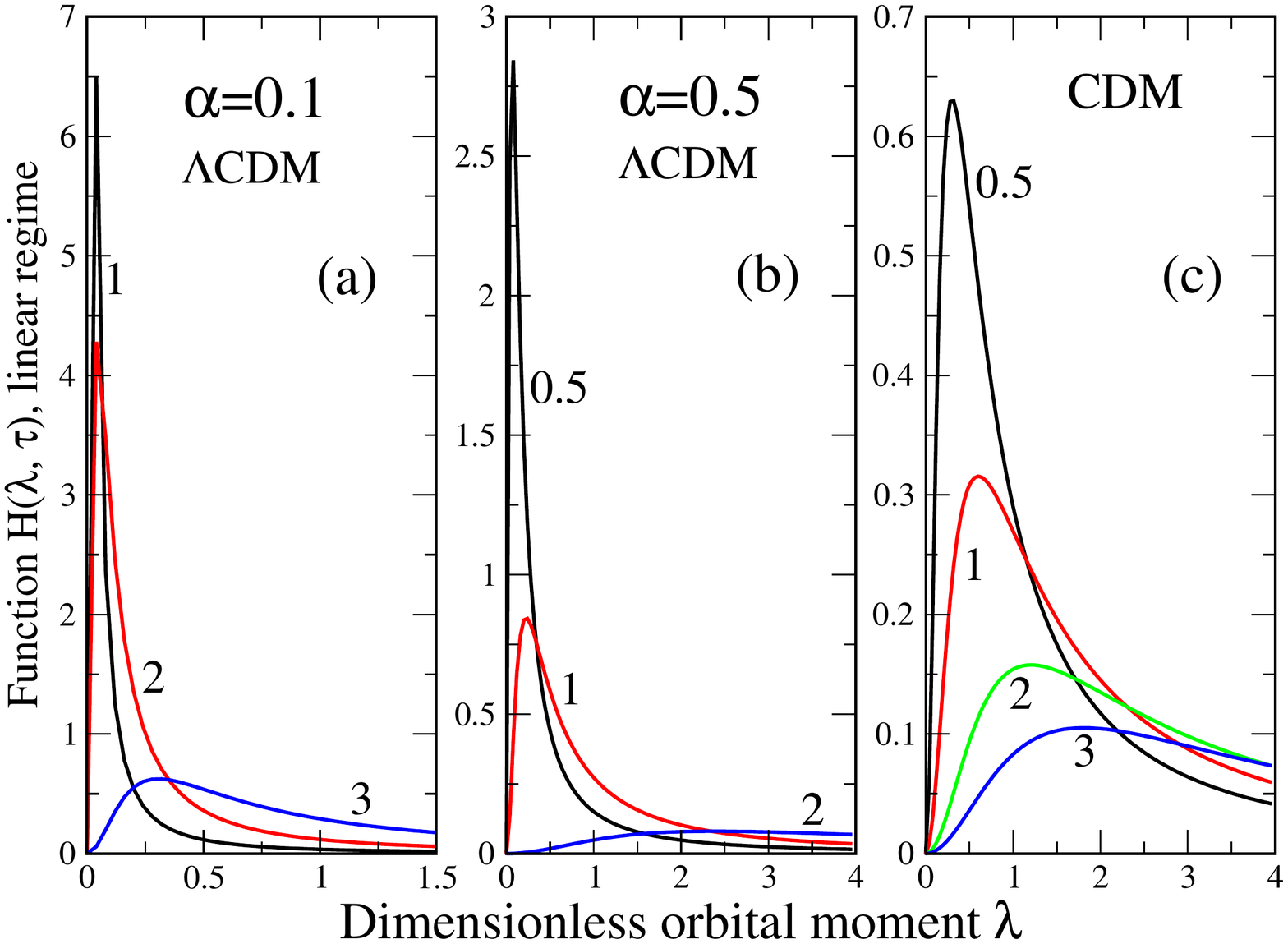}
\includegraphics[width=0.8\columnwidth]{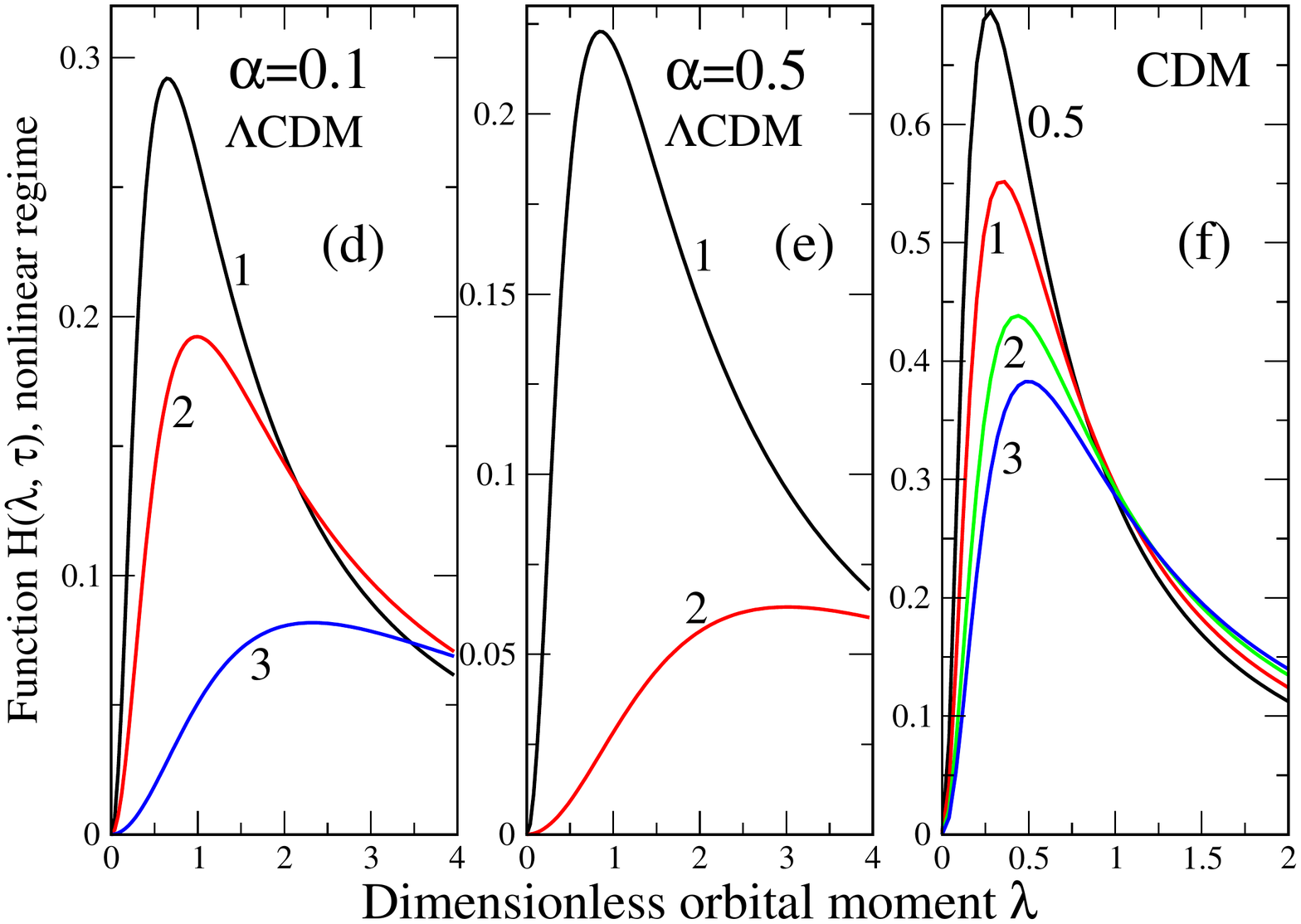}
\caption{One dimensional effective distribution function $H(\lambda,\tau)$. The figure reports time evolution of above 
function in both $\Lambda$CDM and CDM (panels (c) and (f)) models, see legends. We present also differences between linear (upper panels) and nonlinear (lower panels) regimes. Figures near curves correspond to dimensionless time $\tau=t/t_0$.}
\label{fig2}
\end{figure}

The above results lead to conclusion that angular momentum of galaxy clusters should increase in time. This hypothesis could be tested theoretically. This is because limited speed of light causes that the age of the astronomical objects with different redshifts $z$ is different. So, assuming that galaxy clusters form in the same time instant, we expect that clusters with higher redshift $z$ are younger. This means that galaxies alignment should decrease 
with $z$. Our preliminary analysis of the sample of 247 Abell cluster shows that in the case of $\chi^2$ and Kolmogorov? - Smirnov tests (\cite{g2011b,Aryal13}), the analysed statistics decreases with $z$. Unfortunately this effect is not significant since parameter $t=a/\sigma(a)$ is less then 1. 

One should note however, that basic catalog of galaxies is complete up to magnitude $m=18.^m3$, which 
means that the red shift of the most distant cluster $z<0.12$. As result it is very difficult to detect such subtle effect
for small cluster sample. Moreover, although the vast majority of clusters do not rotate 
(\cite{regos98, Hwang07}), this is not completely true for all clusters (\cite{Hwang07}). \cite{Hwang07} study the dispersions and velocity gradients
in 899 Abell clusters. They have found possible evidence for rotation in only six of them i.e. less then 1\%.
Latter sample of rotating clusters has been studied by \cite{Aryal13}. The random orientation of galaxies angular momenta vectors in the analysed clusters was found. Similarly, \cite{Aryal17} found no preferred alignments of angular momenta vectors of galaxies in a sample of six dynamically unstable clusters. Presence of such cluster types, even relatively small, could give additional difficulties in the observational investigation of the time evolution of the clusters angular momenta. So, larger sample of the cluster stretched for higher $z$ is necessary to make unambiguous conclusions regarding above effect.

\section{Relation to observational results}

Our calculations demonstrate that although the gravitational interaction between stellar components (including dark matter halos) is of long-range multipole character, the observations (see below) give some confirmations that there is additional short-range (like $\sim \exp(-r/r_c)$ with range $r_c$) interaction. As a result, if the distance $r$ between two objects (say galaxies) is smaller then $r_c$, they are correlated which means that  their orbital moments are aligned. This assumption works for the dense (rich) 
galaxy clusters, which, by this virtue, have high degree of orbital moments alignment. For the sparse (poor) clusters the situation is opposite. For such type of clusters  the intergalaxy distance $r>r_c$, the long-range multipole interaction prevails so that there is no alignment of the orbital moments. The above statistical method accounts for this situation if we add the (empirical) short-range interaction term to the initial potential \eqref{star3}. In the analysed case we obtain that the distribution function of random fields would depend on the average angular momentum $L_{max}\equiv L_{av}$ (see \cite{stef97, semstef03}) and as result we obtain the self-consistent equation for $L_{av}$
\begin{equation}\label{lmax}
L_{av}=\int L(E) f(E,L_{av})d^3E.
\end{equation}
where $f(E,L_{av})$ is the distribution function of gravitational field $E$, depending on $L_{av}$ as parameter. This function substitutes the expression \eqref{star19} in the case of inclusion of the possible short-range interaction term. One should note that in the case of finite $r_c$, the distribution 
function decays at $E \to \infty$ faster then \eqref{star19} so that the integral \eqref{lmax} converges. As  total interaction potential contain both luminous and dark matter components, the equation \eqref{lmax} allows us to ask the question about alignment of sub-dominant galaxies, 	
even though the majority of cluster angular momentum is related to the smooth dark matter halo component. For instance, in the halo model (\cite{sb10}), when the luminous matter of galaxies is embedded in dark matter halo, this halo by virtue of its mass may mediate the intergalaxy interaction, adding possible short-range terms to it. The self-consistent equation \eqref{lmax} permits also to include the temperature into 
consideration (\cite{semstef02,semstef03}) and study the galaxies and their clusters (with respect to dark matter haloes) time evolution within $\Lambda$CDM model. Also, the combination of stochastic models (\cite{gs1,gs2}) of primordial dynamics along with those of $\Lambda$CDM, most probably, would permit to answer (at least theoretically) the question if the galaxies are initially aligned at the time of their formation, or such alignment is generated in some merger events, and how dark matter haloes influence (mediate) this alignment. 

Here we also show that there are different possible relations between angular momentum and the mass (richness) of the cluster. {{Note that $M^{5/3}$ - law for such dependence as well as reasonable values of parameter $\lambda$ in Eq. \eqref{star27a} had been obtained by \cite{heav88} followed by \cite{sch}.}}
Figure \ref{gala} reports our preliminary results of the dependence between analysed statistics obtained for the sample of $247$ rich Abell clusters (\cite{g10a}). We conclude here that our comparison of the cases when the statistics grows as $N$, $N^{4/3}$ and $N^{5/3}$ does not permit to 
establish unambiguous correspondence of different dependences between angular momentum and richness of the structure. However, such unambiguous discrimination would be possible if larger statistical sample of galaxy clusters is available. Moreover, we show that angular momentum of galaxies should increase with time. Latter fact follows from equations \eqref{lm2} - \eqref{lm2b} for CDM model and from Fig. \ref{fig2} for $\Lambda$CDM model. The physical mechanism of that has been discussed in details in our previous paper (\cite{myapj}). It is related to the growing time evolution of scale factor $a(t)$ both in CDM (\cite{dor70}) and $\Lambda$CDM models, see Eq. \eqref{frid3} for details.
This means that above theoretically predicted effect could be tested by observations as galaxies angular momentum should decrease with redshift $z$. Once more, the enlarged sample containing clusters with much higher $z$ is necessary for such studies.

\section{Conclusions}
To summarise, in the present paper we analyze theoretically the observational dependences of the galaxies and their clusters angular momenta on their mass (richness).  To do so, we use the method, introduced in our previous paper (\cite{myapj}). Observational data are in agreement with our theoretical results and mainly  Eqs. \eqref{lm3} and \eqref{lm2a} where we have shown that under reasonable assumptions about cluster morphology the angular momentum of galaxy structures increases with their richness. The solution of equation \eqref{lmax} will permit to establish a relation between the characteristics of possible short-range intergalaxy interaction and character of their spins alignment.

We emphasize however, that the above observational results about lack of alignment of 
galaxies for poor clusters, as well as evidence for such alignment in the rich galaxy clusters 
(\cite{Godlowski05,Aryal07},  see also \cite{g2011a} for later improved analysis) clearly shows that angular momentum of galaxy groups and clusters increases with their richness. The problem of clusters angular momenta in context of their mutual interactions as well as those with dark matter haloes has been discussed by \cite{Hahn07} based on the results of computer simulations. The presence of threshold value of the cluster mass (that is to say richness) has been noticed in these simulations. This threshold value is related to mutual alignment of clusters and dark matter haloes axes. As we have shown above, this fact can be explained by our model.

We finally note that the direct computer simulations of stellar ensembles are still quite computationally expensive to simulate realistic (i.e. sufficiently large) parts of the Universe. Hence it seems to be a good idea to put some effort into developing new theoretical models for galaxy alignment with respect to dark matter haloes and (possible) merger into larger structures like superclusters. Since galaxy morphology plays important role in this behavior, our approach, linking the galaxy shapes with their characteristics distribution (especially in view that it permits to calculate the non-Gaussian pdfs), will improve the overall understanding, which can additionally be tested against observed galaxy shape distributions and alignments.

\appendix
\section{}

Here we present some more details of our model, based on Hamiltonian \eqref{star1}. In this Hamiltonian, the 
explicit expression for $i$ - th galaxy quadrupolar moment $Q_i$ has the form (\cite{poisson98})
\begin{equation}\label{stam2}
Q_i=\int_{V_i} \rho_i({\bf x})|{\bf x}|^2P_2({\bf s}\cdot{\bf x})d^3x,
\end{equation}
where $P_2(x)=(3x^2-1)/2$ is corresponding Legendre polynomial (\cite{abr}), $V_i$ is a volume of $i$-th galaxy, 
$\rho_i({\bf x})$ is a density of its mass.

\begin{figure}
\includegraphics[width=0.7\columnwidth]{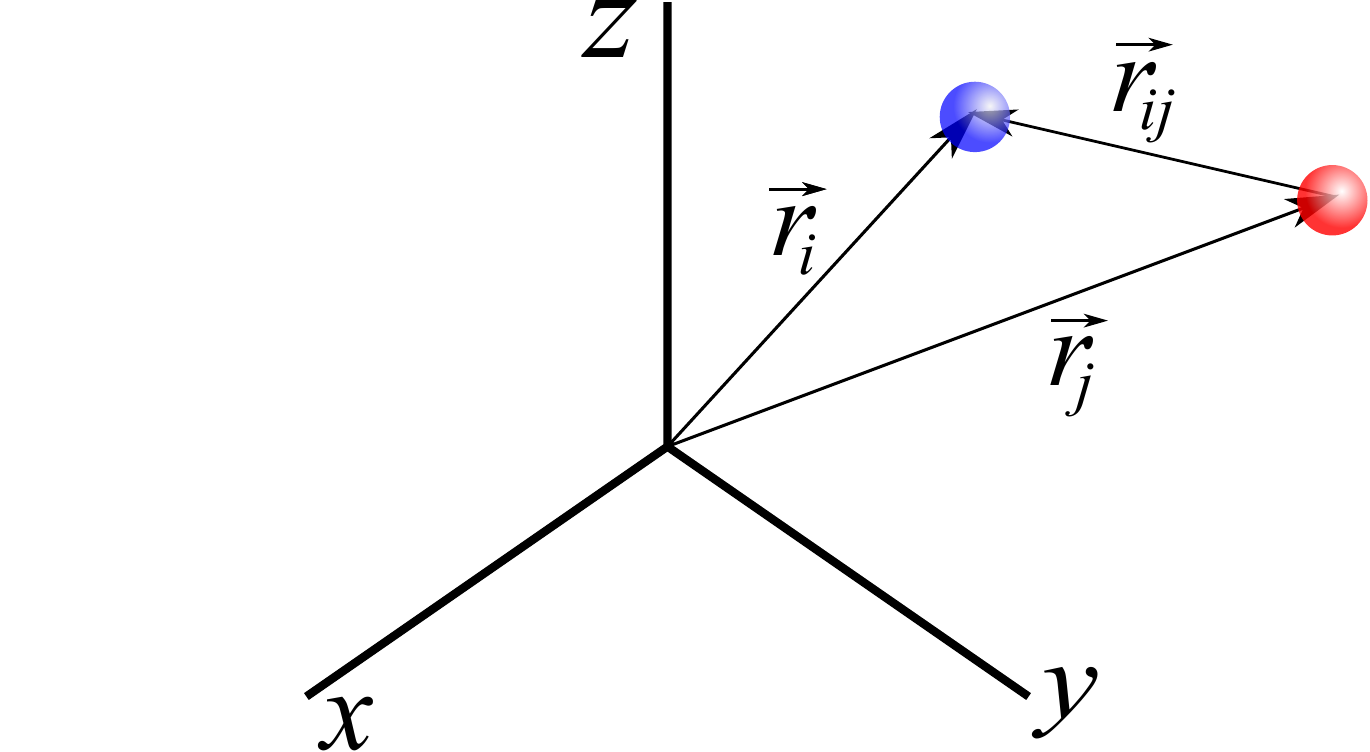}
\caption{The reference frame of the problem under consideration. Radius - vectors of galaxy (or dark matter 
halo element) $i$ (blue ball) and $j$ (red ball) (${\bf r}_i$ and ${\bf r}_j$ respectively) as well as their difference
${\bf r}_{ij}$ are shown.}
\label{fd1}
\end{figure}

The geometry of the problem under consideration is shown in Fig.\ref{fd1}. It is seen first, that the origin is 
not related to any specific galaxy or other astronomical object. Rather, it is situated in the arbitrary point 
in the Universe. Although ${\bf r}_{ij}$ is directed from one galaxy (in our case $j$) towards another (in our 
case $i$) it is by no means bounded to these galaxies. It is simply means the difference in their radius - vectors, 
which connect the coordinates origin and position of each galaxy.

\begin{figure}
\begin{center}
\includegraphics[width=0.8\columnwidth]{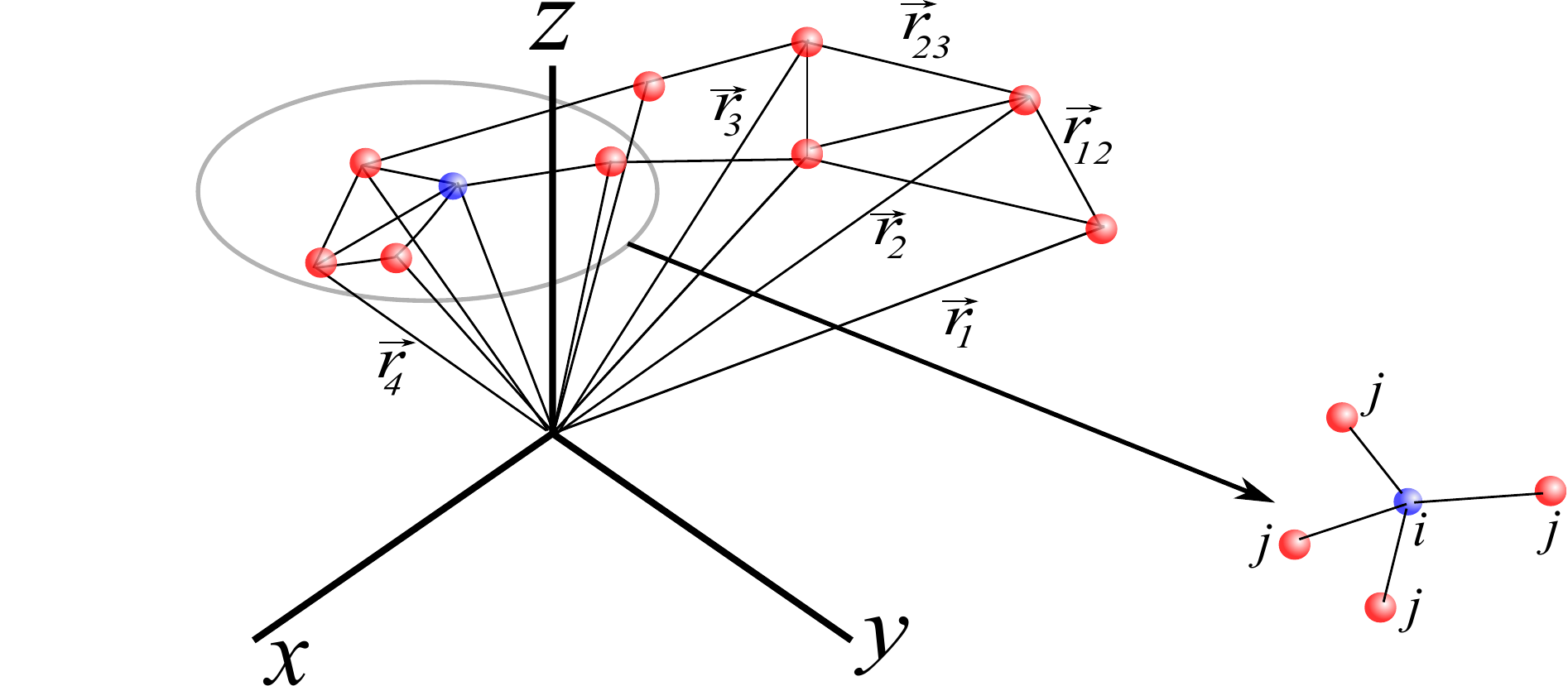}
\end{center}
\caption{Geometry of the problem with many galaxies (or other astronomical objects marked by red and blue balls) 
situated randomly in the Universe. Radius - vectors of those elements (like ${\bf r}_1$, ${\bf r}_2$ etc) as well 
as their separations (like ${\bf r}_{23}$) are shown selectively. Blue ball (in the ellipse on the main panel and 
in the inset) shows the example of $i$-th object with the rest being $j$-th objects. Division on $i$ and $j$ objects 
is arbitrary and made to calculate the gravitational field, exerted on $i$-th object from the rest of the ensemble. 
In other words, any galaxy can be either of $i$ or $j$ type. Inset shows this situation (from the ellipse on the 
main panel): the gravitational field on the (arbitrary chosen) blue ball $i$ is a sum of the fields from its neighboring 
objects $j$. The dimensions of the ellipse on the main panel visualize the range of interaction \eqref{sta3}; this range 
is very long (decays as $r^{-4}$ so that much more galaxies will be in the range of interaction, but the distant $j$-th galaxies make almost zero contribution to the gravity field on $i$-th one), it does not have clear boundary but the ellipse gives some 
guide for eyes. As the number of galaxies is actually infinite and their separations become progressively smaller, the galaxies 
connecting polyline (i.e. line consisting of all ${\bf r}_{ij}$) tends to continuous curve (not shown). In this case all sums 
are converted to integrals, as described in the text.}  
\label{fg2}
\end{figure}

The Hamiltonian \eqref{star1} can be identically rewritten through the interaction energy
\begin{eqnarray}
{\cal H}=-GM^2\sum_{i}p_im_iW_i,  \label{sta2} \\
W_i=W({\bf r}_{i})=\sum_j m_jV({\bf r}_{ij})\equiv \nonumber \\
\sum_j m_jV({\bf r}_{j}-{\bf r}_{i}).\nonumber 
\end{eqnarray}
The interaction energy $W_i$ is the energy exerted by the rest of the galaxy ensemble (due to intergalaxy interaction) to 
the galaxy in the point $i$. We can see that after summation (actually integration, see below) over 
${\bf r}_{j}$ the relative intergalaxy distance ${\bf r}_{ij}$ has actually disappeared.  

The gradient of the energy \eqref{sta2} is indeed the gravity field,
which acts on $i$-th galaxy (or other astronomical object) from the rest $j$ of these objects ensemble
 \begin{eqnarray} 
{\bf E}_{quad}({\bf r}_{i})\equiv {\bf E}_{quad,i}=\sum_j m_j\nabla V({\bf r}_{j}-{\bf r}_{i})=\nonumber \\ 
{\bf i}_rE_0\sum_jm_j\frac{3\cos^2\theta_{ij} - 1}{r_{ij}^4}, \label{sta3}
\end{eqnarray}
which is the expression \eqref{star3} from the text, rewritten explicitly in terms of vectors ${\bf r}_{i}$ and 
${\bf r}_j$.  

Having the expression \eqref{sta3}, we can write explicitly the distribution function of random quadrupolar fields,
Eq. \eqref{star8} from the text
\begin{eqnarray}\label{sta4}
f({\bf E})=\overline{\delta({\bf E}-{\bf E}_i)}\equiv \overline{\delta({\bf E}-{\bf E}_{quad}({\bf r}_{i}))}=\nonumber \\
\nonumber \\
=\overline{\delta\left({\bf E}-{\bf i}_rE_0\sum_jm_j\frac{3\cos^2\theta_{ij} - 1}{r_{ij}^4}\right)},
\end{eqnarray}
where bar means the averaging over random spatial configurations of galaxies and other astronomical objects.

In performing the actual averagings in the expression \eqref{sta4} (see Fig.\ref{fg2}), with respect to the fact that number of 
galaxies is infinite and their "elementary separations" ${\bf r}_{ij}$ become very small, we can change summations in \eqref{sta3} 
and \eqref{sta4} to integrations using the expression for gravity field ${\bf E}_i$ in the form \eqref{star3}. Further averagings in \eqref{sta4} are prescribed in the text, see also \cite{myapj}.


\end{document}